\documentclass{amsart}
\usepackage[latin2]{inputenc}
\usepackage{amssymb,amsthm,euscript,amsmath,graphicx}

\newcommand{\be}{\begin{equation}}
\newcommand{\ee}{\end{equation}}
\newcommand{\ben}{\begin{equation*}}
\newcommand{\een}{\end{equation*}}
\newcommand{\bga}{\begin{gather}}
\newcommand{\ega}{\end{gather}}
\newcommand{\bea}{\begin{eqnarray}}
\newcommand{\eea}{\end{eqnarray}}
\newcommand{\eean}{\end{eqnarray*}}
\newcommand{\bean}{\begin{eqnarray*}}
\newcommand\re[1]{(\ref{#1})}

\newcommand\al{\alpha}

\newcommand\ga{\gamma}

\begin{document}
\title{Internal energy in dissipative relativistic fluids}

\author{P\'eter V\'an}
\address{KFKI Research Institute for Particle and Nuclear Physics
 Budapest and\\
 BCCS Bergen Computational Physics Laboratory, Bergen
}

\email{vpet@rmki.kfki.hu}

\date{\today}

\begin{abstract}
Liu procedure is applied to a first order weakly nonlocal special
relativistic fluid. It is shown, that a reasonable relativistic
theory is and extended one, where the basic state space contains the
momentum density. This property follows from the structure of the
energy-momentum balance and the Second Law of thermodynamics.
Moreover, the entropy depends on the energy density and the momentum
density on a given specific way, indicating that the local rest
frame energy density cannot be interpreted as the internal energy,
the local rest frame momentum density should be considered, too. The
corresponding constitutive relations for the stress and the energy
flux are derived.
\end{abstract}
\maketitle


\section{Introduction}

Nonrelativistic nonequilibrium thermodynamics separates the
dissipative and nondissipative parts of the evolution of physical
quantities. This separation is based on the construction of the
internal energy balance \cite{Eck40a1,GroMaz62b,Gya70b}. According
to the classical interpretation the internal energy is the
difference of the total energy and the known specific energy types.
The entropy function depends directly on the internal energy. The
internal energy is distributed unbiased among the molecular degrees
of freedom. The process how the other energy types are converted to
internal energy is called dissipation. This approach is common in
every theories of nonequilibrium thermodynamics including classical
irreversible thermodynamics, where the hypothesis of local
equilibrium applies. However, there is nothing similar in
relativistic irreversible thermodynamics. Furthermore, practically
there is no relativistic irreversible thermodynamics at all, because
the local equilibrium theory is plagued by serious controversies,
therefore only the extended theories, theories beyond local
equilibrium, are considered as viable. The reason is that the
classical theory of Eckart for relativistic fluids \cite{Eck40a3} is
simple and elegant, but produces generic instabilities
\cite{HisLin85a}. The more developed extended theories incorporate
the theory of Eckart, therefore inherit (but more or less suppress)
the instabilities \cite{HisLin87a,Ger95a,Lind96a}.

In this paper we investigate the possibility of local equilibrium in
relativistic hydrodynamics by methods of continuum thermodynamics.
At the next section the balances of energy-momentum and entropy are
introduced. In the third section we calculate the dissipation
inequality for local equilibrium (first order) relativistic
hydrodynamics by Liu procedure. The need of second order (extended,
or weakly nonlocal) theories is indicated by the emergent structure.
A new concept of relativistic internal energy follows. Based on
these results we give the constitutive equations of the simplest
extended theory by the heuristic arguments of irreversible
thermodynamics in the fourth section.

\section{Basic balances of relativistic fluids}

For the metric (Lorentz form) we use the $g^{\mu\nu}=
diag(-1,1,1,1)$ convention and the units are introduced that the
speed of light $c=1$. Therefore for a four-velocity $u^\al$ we have
$u_\al u^\al = -1$. $\Delta^\al_{\;\;\beta} = g^\al_{\;\;\beta} +
u^\al u_\beta$ denotes the $u$-orthogonal projection. First we give
the basic balances of energy-momentum and entropy.

The energy-momentum density tensor is given with the help of the
rest-frame quantities
 \be
  T^{\alpha\beta} = e u^\alpha u^\beta + u^\alpha q^\beta +
    u^\beta {q}^\alpha + P^{\alpha\beta},
 \label{T}\ee

\noindent where $e = u_\alpha u_\beta T^{\alpha\beta}$ is the {\em
density of the energy}, $q^\beta = -
u_\alpha\Delta^\beta_{\;\;\gamma} T^{\alpha\gamma}$ is the  {\em
energy flux} or {\em heat flow} ${q}^\alpha = -
u_\beta\Delta^\alpha_{\;\;\gamma} T^{\gamma\beta}$ is the {\em
momentum density} and $P^{\alpha\beta}= \Delta^\alpha_\gamma
\Delta^\beta_\mu T^{\gamma\mu}$ is the {\em pressure tensor}. The
momentum density, the energy flux and the pressure are spacelike in
the comoving frame, therefore $u_\alpha {q}^\alpha = 0$ and
$u_\alpha q^\alpha = 0$ and $u_\al P^{\alpha\beta} = u_\al
P^{\beta\al} = 0^\beta$. The energy-momentum tensor is symmetric,
because we assume that the internal spin of the material is zero. In
this case the energy flux and the momentum density are equal.  Let
us emphasize that the form \re{T} of the symmetric energy-momentum
tensor is completely general for one-component fluids, nevertheless
it is expressed by the local rest frame quantities.

Now the conservation of energy-momentum  $\partial_\beta
T^{\alpha\beta} = 0$ is expanded to \be
 \partial_\beta T^{\alpha\beta} =
    \dot{e} u^\al + e u^\al \partial_\beta u^\beta +e\dot{u}^\al +
    u^\al\partial_\beta q^\beta +q^\beta \partial_\beta u^\al +
    \dot{{q}}^\al + {q}^\al \partial_\beta u^\beta +
    \partial_\beta P^{\al\beta},
\label{E-Ibal}\ee

\noindent where $\dot{e} = \frac{d}{d\tau}e = u^\alpha
\partial_\alpha e$ denotes the derivative of $e$ by the proper time
$\tau$. Its timelike part in a local rest frame gives the balance of
the energy
 \be
    -u_\alpha\partial_\beta T^{\alpha\beta} =
        \dot{e} + e \partial_\alpha u^\alpha +
        \partial_\alpha q^\alpha +
        {q}^\alpha \dot{u}_\alpha +
        P^{\al\beta} \partial_\beta u_\alpha = 0.
 \label{ebal}\ee

The spacelike part in the local rest frame describes the balance of
the momentum
 \be
    \Delta^\alpha_{\;\;\gamma}\partial_\beta T^{\gamma\beta} =
        e\dot{u}^\alpha  +
        {q}^\alpha \partial_\beta u^\beta +
        q^\beta \partial_\beta u^\alpha +
        \Delta^\alpha_{\;\;\gamma} \dot{{q}}^\gamma +
        \Delta^\al_{\;\;\ga}\partial_\beta P^{\ga\beta}
    = 0^\al.
 \label{ibal}\ee

The entropy density and flux can also be combined into a
four-vector, using local rest frame quantities:
 \be
  S^\alpha = s u^\alpha + J^\alpha,
 \label{S}\ee
\noindent where $s= -u_\alpha S^\alpha$ is the  {\em entropy
density} and $J^\alpha = S^\alpha - u^\alpha s =
\Delta^\alpha_{\;\;\beta} S^\beta$ is the {\em entropy flux}. The
entropy flux is $u$-spacelike, therefore $u_\alpha J^\alpha = 0$.
Now the Second Law of thermodynamics is expressed by the following
inequality
 \be
 \partial_\alpha S^\alpha = \dot{s} + s\partial_\alpha u^\alpha +
    \partial_\alpha J^\alpha \geq 0.
 \label{sbal}\ee

\section{Thermodynamics}

The thermodynamical background in relativistic theories is usually
based on analogies with nonrelativistic thermostatics. However,
nonequilibrium thermodynamics developed beyond the simple 'let us
substitute everything into the entropy balance and see what happens'
theory since Eckart. It is important to check the dynamic
consistency of the Second Law, considering the evolution equations
as constraints for the entropy balance. This method of
nonequilibrium thermodynamics is constructive, gives important
information for new theories and reveals some deeper interrelations.
Here we exploit the Second Law by Liu's procedure \cite{Liu72a}
introducing a first order weakly nonlocal state space in all basic
variables, thus restricting ourselves to a local equilibrium theory.
One can find a general treatment of nonrelativistic classical and
extended irreversible thermodynamics from this point of view in
\cite{Van03a}. Our aim here is to investigate the relativistic
fluids with similar methods to get the relativistic equivalent of
the classical Fourier-Navier-Stokes system of equations for one
component fluids.

Our most important assumption regarding relativistic thermodynamics
is that the constitutive equations are local rest frame expressions.
From a physical point of view it is natural, because material
interactions are local.

The {\em basic state space} of the theory is spanned by the energy
density $e$ and by the velocity field $u^\al$. The {\em constitutive
state space} is spanned by the basic state variables and their first
derivatives, therefore it is first order weakly nonlocal. Hence the
constitutive functions depend on the variable set $C=(e, u_\al,
\partial_\al e, \partial_\al u_\beta)$. The {\em constitutive
functions} are the energy flux/momentum density $q^\al$, the
pressure $P^{\al\beta}$, the entropy density $s$ and the entropy
flux $J^\al$. The derivatives of the constitutive functions are
denoted by the serial number of the corresponding variable in the
constitutive space, e.g. $\frac{\partial s}{\partial (\partial_\al
e)} = \partial_3 s$. With this notation we can distinguish easily
the derivatives by the constitutive and spacetime variables. A
nonequilibrium thermodynamic theory is considered to be solved if
all other constitutive quantities are expressed by the entropy
density and its derivatives.

According to the procedure of Liu the balance of energy-momentum
\re{E-Ibal} is a constraint to the entropy balance \re{sbal} with
the Lagrange-Farkas multiplier $\Lambda_\al$
 \be
 \partial_\al S^\al - \Lambda_\al\partial_\beta T^{\al\beta}\geq 0.
\label{sl1}\ee

Let us remember, that here the spacelike components of the four
quantities and the entropy density are the constitutive quantities
depending on the introduced constitutive variables $C$. Therefore,
in the above inequality we can develop the derivatives of the
composite functions. The coefficients of the derivatives that are
not in the constitutive space must be zero, therefore we get the
following Liu-equations:
 \bea
   \partial_{\al\beta} e &:&
     (\partial_{3} S^\al)^{\beta} -
     \Lambda_\mu (\partial_{3}T^{\mu\al})^\beta = 0^{\al\beta},
            \label{liu1}\\
   \partial_{\al\beta} u_\ga &:&
        (\partial_{4} S^\al)^{\beta\ga} -
            \Lambda_\mu(\partial_{4} T^{\mu\al})^{\beta\ga}
        = 0^{\al\beta\ga}. \label{liu2}
\eea

The simple structure of the Liu equations suggests the assumption
that the Lagrange multiplier is a local function, it does not depend
on the derivatives of the basic state variables
 \be
 \Lambda_\ga = \Lambda_\ga(n, e).
\label{ass1}\ee

Then a general solution of \re{liu1}-\re{liu2} is
 \be
  S^\al -\Lambda_\ga {T}^{\ga\al} - A^\alpha =0^\alpha,
\label{lmo} \ee

\noindent where $A^\al = A^\al (n,e)$ is an arbitrary local
function.

Let us introduce the splitting of the vector multiplier and the
four-vector $A^\al$ to spacelike and timelike parts in the local
rest frame as
 \bean
    \Lambda^\al &=& -\Lambda u^\al + l^\al, \\
    A^\al       &=& A u^\al + a^\al,
\eean

\noindent where for the spacelike components $ u_\al l^\al= u_\al
a^\al =0$. Now equation \re{lmo} gives
 \be
  u^\al(s - \Lambda e - l_\ga{q}^\ga -A) +
    (J^\al - \Lambda {q}^\al -  l_\ga P^{\ga\al} -
        a^\al) = 0^\al.
\label{llc}\ee

Here both the timelike and spacelike parts are zero, resulting in
 \bea
  s &=& \Lambda e + l_\ga{q}^\ga +A, \label{GDf}\\
  J^\al &=& \Lambda {q}^\al + l_\ga P^{\ga\al} + a^\al.
 \label{ef}\eea

After the identification of the Liu equations we can get the {\em
dissipation inequality} as
\begin{gather}
 \partial_\al e \left[
    (\partial_1 s) u^\al +
    \partial_1 J^\al -
    \Lambda u^\al -
    \Lambda \partial_1 q^\al -
    l_\ga \partial_1 P^{\ga\al}-
    l_\ga\partial_1 {q}^\ga u^\al  \right]  +
        \nonumber\\
 \partial_\al u_\beta \left[
    \left(s- \Lambda e -l_\ga{q}^\ga \right) \Delta^{\al\beta} +
    (\partial_2 s)^\beta u^\al +
    (\partial_2 J^\al)^\beta -
     \right.\nonumber\\\left.
     l^\beta e u^\al -
    l^\beta q^\al -
    \Lambda (\partial_2 {q}^\beta)^\al  -
    \Lambda_\ga (\partial_2 P^{\ga\al})^\beta -
    \Lambda_\ga u^\al (\partial_2 {q}^\ga)^\beta
    \right]\geq 0.
\label{di1}\end{gather}

Here we exploited that the partial differentiation by $e$ can be
exchanged with the multiplication by the four velocity $u^\al$.

In the dissipation inequality one should consider the solution of
the Liu-equations. Substituting \re{GDf} and \re{ef} into \re{di1}
we get
 \begin{gather}
 \partial_\al e\left[
    (\partial_1 s -\Lambda -
        l_\ga \partial_1 {q}^\ga) u^\al +
    q^\al \partial_1 \Lambda +
    P^{\ga\al} \partial_1 l_\ga +
    \partial_1 a^\al\right]  +\nonumber\\
 \partial_\al u_\beta\left[
    A\Delta^{\al\beta} +
    q^\al (\partial_2 \Lambda)^\beta  +
    P^{\ga\al}(\partial_2l_\ga)^\beta +
    (\partial_2a^\al)^\beta  + \right. \nonumber\\\left.\quad
    u^\al\left((\partial_2 s)^\beta -
        l_\ga(\partial_2 q^\ga)^\beta -
        l^\beta e -
        \Lambda q^\beta\right) -
    l^\beta q^\al -
    \Lambda P^{\al\beta}
    \right]  \geq 0. 
\end{gather}

Here the following identities were applied simplifying the last term
($\partial_2 = \partial_{u_\beta}$)
 \bean
 u_\ga\partial_{u_\beta}q^\ga &=& \partial_{u_\beta}(u_\ga q^\ga) -
    q^\ga\partial_{u_\beta}u_\ga = -q^\ga \Delta_\ga^{\;\;\beta} =
    -q^\beta, \\
 u_\ga\partial_{u_\beta}P^{\ga\al} &=&
    \partial_{u_\beta}(u_\ga P^{\ga\al}) -
        P^{\ga\al}\partial_{u_\beta}u_\ga =
    -P^{\ga\al}\Delta_\ga^{\;\;\beta} =
    -P^{\beta\al}.
\eean

Observing the first term in the last form of the dissipation
inequality one can eliminate the direct velocity dependence of the
entropy function recognizing that the entropy may depend on the
energy flux in the following form
 \be
    s(e,u^\al, \partial_\al e, \partial_\al u^\beta)=
    \hat{s}(e,q^\ga(e,u^\al,\partial_\al e, \partial_\al u^\beta)).
 \label{lEnt}\ee

Therefore the entropy is local, independent of the derivatives of
the basic state space variables and the velocity field, but depends
on the energy flux. The energy flux can depend also on the
derivatives because according to our initial assumptions it is
considered as a constitutive function. Then the Lagrange-Farkas
multipliers are determined by the entropy derivatives
 \be
  \partial_e \hat{s} = \Lambda, \quad
  \partial_{q^\al} \hat{s} = l_\al.
 \label{intd}\ee

We introduce a temperature $T$ as
 \be
   \partial_e \hat{s} = \Lambda = \frac{1}{T}.
 \label{Tem}\ee
We may recognize that a full thermostatic compatibility requires
that in \re{GDf}: $A :=\frac{p}{T}$, where $p$ is the pressure.
These consequences are completely analogous to the results of the
nonrelativistic nonequilibrium thermodynamic theory, where
thermostatics arises from the structure of the balance form
evolution equations as constraints for the Second Law.

Finally we assume that entropy flux is classical and the additional
term $a^\al$ \cite{Mul67a} is zero
 \be
   a^\al = 0^\al.
\label{ass3}\ee

Then the dissipation inequality reduces to the following simple form
 \be
  q^\al \partial_\al \frac{1}{T} -
  \frac{1}{T} \left( P^{\al\beta} +
    T l^\beta q^\al-
    p\Delta^{\al\beta}\right)\partial_\al u_\beta -
  P^{\al\ga}\partial_\al l_\ga -
  \left(\frac{q^\al}{T} + e l^\al \right)\dot{u}_\al \geq 0.
 \label{di3}\ee

As we do not want an acceleration dependent entropy production, we
must require that the last term vanishes. According to \re{intd} and
\re{Tem}
 \be
  e \partial_{q_\al} \hat{s} + q^\al \partial_e \hat{s}= 0.
 \label{pdfe}\ee

The general solution of  \re{pdfe} can be given as
 \be
 \hat{s}=\tilde{s}(e^2 - q^\al q_\al) + B,
 \label{sform}\ee

\noindent where $B$=const. The entropy must depend on the energy
density $e$ and the momentum density $q^\al$ on a very particular
but simple way. As a consequence of this functional form of the
entropy function the Gibbs relation can be given with the help of
the entropy derivatives \re{intd} as
 \be
 de - \frac{q^\al}{e} dq_\al = T ds.
\label{thermo}\ee

We may require first order homogeneity of the entropy in \re{sform},
without restricting the generality. That can we get introducing $E =
\sqrt{|e^2-q_\al q^\al|}$ as a variable of the entropy density. In
this way the entropy is a first order homogeneous functions both of
the energy density $e$ and the momentum density $q^\al$. With this
property it is unique.

The corresponding potential relation can be constructed according to
the first order homogeneity of the physical quantities as
 \be
 e-\frac{q^\al q_\al}{e} = T s - p.
\label{potrel}\ee

The previous thermostatic relations require the interpretation of
$E$ as internal energy. On the other hand let us recognize that $E$
is the absolute value of the energy vector
 \be
  E= \|E^\al\| = \|-u_\beta T^{\beta\al}\|=\| e u^\al + q^\al\|=
  \sqrt{|e^2-q_\al q^\al|}.
  \ee

However, one should pay attention that the $1/T$ introduced in
\re{Tem} is not the derivative of the entropy function according to
$E$.

Finally the entropy flux from \re{ef} and \re{ass3}
 \be
  J^\al = \frac{1}{T} {q}^\al  -\frac{q_\gamma}{e T} P^{\ga\al}.
\label{ref}\ee

The final form of the dissipation inequality is
 \be
  q^\al \partial_\al \frac{1}{T} -
  \frac{1}{T} \left( P^{\al\beta} +
    \frac{q^\beta q^\al}{e}-
    p\Delta^{\al\beta}\right)\partial_\al u_\beta -
  P^{\al\ga}\partial_\al \frac{q_\ga}{T e} \geq 0.
 \label{di4}\ee

The last term in this expression with a derivative of one of the
constitutive quantities indicates that we cannot give proper
thermodynamic fluxes and forces, as a solution of the inequality. An
other problem appeared already with \re{pdfe}, because $l_\al$, as
the spacelike part of the Lagrange multiplier in a local rest frame,
was assumed independent of the derivatives of $e$ and $u^\al$. That
is the Fourier heat conduction was excluded as a possible
constitutive function. Both problems indicate that a complete theory
may exist only either in an enlarged constitutive space or in an
extended basic state space. One of the possibilities is to introduce
higher order derivatives of the basic state space into the
constitutive state space and construct a second order weakly
nonlocal theory. The other possibility is to enlarge the basic state
space and construct an extended theory. In both cases the key that
may lead beyond the traditional Müller-Israel-Stewart theory is the
new internal energy $E$.

\section{Extended irreversible thermodynamics of relativistic fluids}

Motivated by the results of the previous section we calculate the
entropy production by a direct substitution of the balance of the
energy into the entropy balance. We are to construct an extended
theory, introducing $q^\al$ as an independent variable, but
exploiting the fact that the entropy depends both on the energy and
momentum densities by the specific way derived above.

The entropy flux is assumed to have the very classical form \be
 J^\alpha = \frac{1}{T }q^\al.
 \label{sf}\ee

Substituting the energy balance \re{ebal} into the entropy balance
equation we arrive at the following entropy production formula:
 \begin{gather}
  \partial_\alpha S^\alpha =
    \dot{s}(e^2+ q^\al q_\al, s) + s \partial_\alpha u^\alpha +
        \partial_\alpha J^\al = \nonumber \\
     -\frac{1}{T }(e\partial_\alpha u^\alpha +
       \partial_\alpha q^\alpha + q^\al \dot{u}_\al +
        P^{\al\beta}\partial_\beta u_\al) +
    \frac{q^\al}{T e}\dot{q}_\al +
    s\partial_\alpha u^\alpha +
    \partial_\alpha \left( \frac{1}{T }{q}^\alpha \right)= \nonumber\\
   -\frac{1}{T }\left(P^{\alpha\beta} -
    (-e+sT)\Delta^{\alpha\beta}\right)
        \partial_\alpha u_\beta  +
    {q}^\alpha\left(
        \partial_\alpha\frac{1}{T}
        -\frac{\dot{u}^\al}{T} - \frac{\dot{q}^\al}{e T}
       \right)\geq 0
 \label{cld}\end{gather}

In isotropic continua the above entropy production results in
constitutive functions assuming a linear relationship between the
thermodynamic fluxes and forces. The thermodynamic fluxes are the
{\em viscous stress} $\Pi^{\al\beta}=\left(P^{\alpha\beta} -
(sT-e)\Delta^{\alpha\beta}\right)$, and the energy flux $q^\al$. We
get
  \bea
   \Pi^{\alpha\beta} &=&
        P^{\alpha\beta} -\Delta^{\alpha\beta}\left(p-
            \frac{q^\beta q_\beta}{e}\right)=
        -2\eta (\Delta^{\alpha\gamma}\Delta^{\beta\mu}\partial_\gamma u_\mu)^{s0} -
        \eta_v \partial_\gamma u^\gamma \Delta^{\alpha\beta}, \label{newt}\\
   q^\alpha &=&
   -\lambda \frac{1}{T^2} \Delta^{\alpha\gamma}
    \left(\partial_\gamma T +
        T \dot{u}^\al +
        \frac{T \dot{q}^\al}{e} \right),
\label{fo} \eea

\noindent where $^{s0}$ denotes symmetric traceless part of the
corresponding second order tensor $(A^{ij})^{s0} =
\frac{1}{2}(A^{ij}+A^{ji}) - \frac{1}{3} A^{ll} \delta^{ij}$ and we
have introduced the scalar thermostatic pressure according to
\re{potrel} (therefore $p\neq P^\al_\al/3$). \re{newt} and \re{fo}
are the relativistic generalizations of the Newtonian viscous stress
function and the Fourier law of heat conduction. The shear and bulk
viscosity coefficients, $\eta$ and $\eta_v$ and the heat conduction
coefficient $\lambda$ are nonnegative, according to the inequality
of the entropy production \re{cld}.

The equations \re{ebal}, \re{ibal} are the evolution equations of a
relativistic heat conducting ideal fluid, together with the
constitutive functions \re{newt} and \re{fo}. As special cases we
can get the relativistic Navier-Stokes equation substituting
\re{newt} into \re{ibal} and assuming $q^\al = 0$, or the
relativistic heat conduction equation substituting \re{fo} into
\re{ebal} assuming that $\Pi^{\al\beta}=0$. The heat conduction part
results in a special extended theory, where only the energy flux
appears as independent variable.

\section{Summary and discussion}

Int the first part of the paper we have investigated the local
equilibrium theory of special relativistic fluids. We have seen that
there may be no such theory that could give a complete solution of
the entropy inequality provided the following conditions
\begin{enumerate}
 \item local Lagrange-Farkas multipliers,
 \item local entropy \re{lEnt},
 \item no additional term in the entropy flux \re{ass3}.
\end{enumerate}

The first two assumptions were necessary to get a particular
solution of the Liu equations and the dissipation inequality. On the
other hand they are natural in local equilibrium.

We have concluded that either an extension of the basic state space
or an enlargement of the constitutive state space can give a
complete solution. Our investigations indicated a particular
dependence of the entropy on the energy and momentum densities,
leading to a distinction of internal and total energy densities of
relativistic fluids.

The local rest frame energy density $e = u_\al T^{\al\beta}u_\beta$
is usually interpreted as internal energy in thermodynamic theories.
However, the symmetry of the energy-momentum tensor can hide that
energy flux is related to dissipation, but momentum density is not.
According to the previous investigations the total energy density
$e$ - the time-timelike part of the energy-momentum tensor - is not
a suitable internal energy, the entropy density should be a function
of the absolute value of the energy vector $E^\al = -u_\beta
T^{\al\beta}$ - the timelike part of the energy momentum.

To compare our proposal to the traditional Müller-Israel-Stewart
theory \cite{Isr76a,IsrSte80a} it is instructive to expand the
internal energy into series, assuming that $e^2 > q^\al q_\al$
 \be
E = \sqrt{|e^2 - q^\al q_\al|} \approx e -
    \frac{\bold q^2}{2 e} + ... .
\label{ie}\ee

The last, quadratic term in the above expression is what appears in
the Müller-Israel-Stewart theory. However, in our case
 \begin{itemize}
\item the corresponding relaxation time is fixed $\tau = 1/e$,
\item the quadratic term is only the first approximation,
\item only the energy flux was introduced as an independent variable
in our extended theory, the viscous stress was not necessary.
\end{itemize}

The series expansion is instructive comparing to nonrelativistic
hydrodynamics. There the internal energy is the difference of the
total energy and the relative kinetic energy. In \re{ie} the
quadratic expression is what one could consider as a kind of energy
of the flow, but of course it is not connected to an external
observer, it considers only the local rest frame momentum density.
In a sense our expression shows that introducing $E$ as internal
energy we declared that the energy of the flow (in the local rest
frame) does not give a dissipative contribution.

The extension of the present calculations considering the balance of
particle number is straightforward. Moreover, one can show that the
above system of equations gives a stable homogeneous equilibrium in
linear stability investigations, contrary to the theory of Eckart
\cite{VanBir07m1}, therefore it can be considered as a minimal
viable extension of the local equilibrium theory without the
complexity of the Müller-Israel-Stewart one.

\section{Acknowledgment}

This work has been supported by the Hungarian National Science Fund
OTKA (T49466, T48489), by the EU-I3HP project and by a Bolyai
scholarship of the Hungarian Academy of Sciences for P.~V\'an.
Enlightening discussions with prof. L\'aszl\'o Csernai are
gratefully acknowledged.

\bibliographystyle{unsrt}

\end{document}